\documentclass{article}%
\usepackage{amsmath}%
\usepackage{amsfonts}%
\usepackage{amssymb}%
\usepackage{graphicx}
\usepackage{subfig}
\usepackage{sidecap}
\usepackage{authblk}
\usepackage{epstopdf}
\usepackage{hyperref}
\usepackage{xcolor}
\usepackage{float}
\usepackage{cite}
\usepackage{url}
\definecolor{dark-red}{rgb}{0.,0.,0}
\definecolor{dark-blue}{rgb}{0.,0.,1}
\definecolor{medium-blue}{rgb}{0,0,1}
\hypersetup{
    colorlinks, linkcolor={dark-red},
    citecolor={dark-blue}, urlcolor={medium-blue}
}

\begin{document}
\normalfont 
\title{Effect of temperature on the effective mass and the neutron skin of nuclei}
\author[1]{E. Y{\"{u}}ksel}
\author[2]{E. Khan}
\author[1]{K. Bozkurt}
\author[3]{G. Col\`{o}}

\affil[1]{Physics Department,Yildiz Technical University, 34220
Esenler, Istanbul, Turkey}
\affil[2]{Institut de Physique Nucl\'eaire, Universit\'e Paris-Sud,
IN2P3-CNRS, F-91406 Orsay Cedex, France}
\affil[3]{Dipartimento di Fisica, Universit\`{a} degli Studi and INFN, Sezione di Milano, 20133 Milano, Italy}

\date{\today}
\maketitle

\begin{abstract}

We study the finite temperature Hartree-Fock-BCS approximation for
selected stable Sn nuclei with zero-range Skyrme forces. Hartree Fock BCS approximation allows for a straightforward
interpretation of the results since it involves $u$ and $v$'s which are not matrices as in HFB.
Pairing transitions from superfluid to the normal state are studied with respect to the
temperature. The temperature dependence of the nuclear radii and
neutron skin are also analyzed. An increase of proton and neutron radii is
obtained in neutron rich nuclei especially above the critical
temperature. 
Using different Skyrme energy functionals, it is found that the correlation 
between the effective mass in symmetric nuclear matter and the critical temperature 
depends on the pairing prescription. The temperature dependence of the nucleon effective mass is also
investigated, showing that proton and neutron effective masses display
different behavior below and above the critical temperature, due to the small temperature dependence of the density.

\end{abstract}

\section{Introduction} 

Pairing correlations in open-shell nuclei are a
crucial part of the mean field calculations which have been
investigated over the years and taken into account within either
Hartree Fock+BCS or HFB approximations at zero temperature. Over the last decades, the effect of temperature on mean field results has
drawn attention due to the vanishing of pairing properties and the phase
change of finite nuclei from superfluid to the normal state at
critical temperatures. Extensions of the mean field models to the
finite temperature case have been given in Refs. \cite{san63,go81,gam13}.
Thereafter, theoretical investigations have been undertaken with
different models and interactions to investigate the relationship
between pairing correlations and temperature. 

First studies in hot finite nuclei were performed with
Hartree-Fock-BCS model \cite{san63,go81}. It has been shown that
pairing correlations disappear and nuclei undergo a sharp transition from
superfluid to normal state at a critical temperature.
Furthermore, the critical temperature and the pairing gap value at zero
temperature follow the $T_{c}\simeq0.567\Delta_{T=0}$ equation
\cite{san63}. In Ref \cite{fan11}, similar calculations have been performed with a correction term in
the nuclear energy density functional. Pairing transitions in finite nuclei were also
predicted within the finite temperature HFB models in Refs.
\cite{go86,egi93,mar03,khan04,mar12}. Recently, the effect of
different types of pairing interactions on the critical temperature
and phase transition of Tin nuclei was investigated within FT-HFB
approximation with zero-range Skyrme interaction. It was shown that
the pairing gap value is sensitive to the type of the pairing force
just below critical temperatures and the critical temperature follows the
$T_{c}\simeq0.5\Delta_{T=0}$ equation \cite{khan07}. Very recently, the
relationship between pairing and temperature effects was also investigated
within the relativistic mean field (RMF) framework in even-even
Ca, Ni, Sn and Pb isotopes in which the critical temperature follows
the $T_{c}=0.6\Delta_{T=0}$ relation \cite{niu13}.
Although the investigation of temperature effect on the pairing
properties of nuclei is limited on the experimental side, some
experimental signatures were also obtained from the level density
measurements of the rare earth $^{161,162}$Dy, $^{171,172}$Yb nuclei.
The observed S-shapes in the semi-experimental heat capacities of
$^{161,162}$Dy, $^{171,172}$Yb \cite{sch01} and $^{166,167}$Er
\cite{mel01} nuclei were interpreted as the phase change of nuclei
from superfluid to the normal state. In addition, the critical
temperature at which the pairing correlations are disappearing was
found at around 0.5 MeV.

Finite temperature
effects are also important for astrophysical events, e.g., the
reaction rates, r-process nucleosynthesis and core-collapse
supernovae. Investigation of the superfluidity and specific heat in
the inner crust matter of neutron stars have also been performed within the
finite temperature HFB framework in Refs. \cite{san04,mon07}. It was shown that
the temperature has an impact on the results and should be included in the
calculations. Furthermore, the effect of temperature on the effective
mass of nucleons, level density and symmetry energy is significant in
the supernovae and electron capture rates which takes place at around
$1<T<2$ MeV \cite{don94,mah85,fan10,fan09,dean02}. Due to the influence of the temperature 
on the proton and neutron effective mass behavior and its connection with astrophysical events,
effect of temperature on nuclei deserves further study. In this framework,
the temperature sensitivity of the effective k-mass
is revealed for the first time in a microscopic approach.

In the present work, fully self consistent calculations are performed within the finite
temperature HFBCS (FT-HFBCS) framework using several types of Skyrme
forces. Since we deal with $u$ and $v$'s in the HFBCS, the interpretation of the temperature effect is more straightforward when compared with HFB approximation. In Section 2, extension of the Hartree Fock BCS model to the
finite temperature case is presented. In Section 3, the phase change
from superfluid to the normal state and effect of temperature on
the neutron skin are analyzed. The relation between the effective mass in symmetric nuclear matter 
and critical temperature is investigated with different Skyrme energy
density functionals. Temperature dependence of the effective nucleon
masses are also analyzed and discussed. Finally, the conclusion is given in Section 4.

\section{Microscopic model: Finite temperature extension of Hartree Fock BCS approximation}

The BCS approach in nuclei is well known \cite{ring80} and extensively
studied over the years. In the Skyrme-HFBCS approach, the expectation value of the Hamiltonian is obtained as:
\begin{equation}
E=\left\langle BCS\left|H'\right|BCS\right\rangle=E[\rho,\kappa,\kappa^{*}]
\end{equation}
and the total energy density functional of nuclei is given by
\begin{equation}\begin{split}
E[\rho,\kappa,\kappa^{*}]&=T+E_{Skyrme}[\rho]+E_{Coul.}[\rho_{p}]+E_{pair}[\rho,\kappa,\kappa^{*}]\\
&=\int d^{3}\textbf{r}[T(\textbf{r})+\mathcal{E}_{Skyrme}(\textbf{r})+\mathcal{E}_{Coul.}(\textbf{r})+\mathcal{E}_{pair}(\textbf{r})].
\end{split}
\end{equation}
Skryme energy density functional for even-even systems is defined as:
\begin{equation}\begin{split}
\mathcal{E}_{Skyrme}(\textbf{r})&=\frac{1}{2}t_{0}\left[\left(1+\frac{x_{0}}{2}\right)\rho^{2}-\left(x_{0}+\frac{1}{2}\right)\sum_{q}\rho_{q}^{2}\right]\\
&+\frac{t_{1}}{4}\left[\left(1+\frac{x_{1}}{2}\right)\left(\rho\tau+\frac{3}{4}(\nabla\rho)^{2}\right)-\left(x_{1}+\frac{1}{2}\right)\sum_{q}\left(\rho_{q}\tau_{q}+\frac{3}{4}(\nabla\rho_{q})^{2}\right)\right]\\
&+\frac{t_{2}}{4}\left[\left(1+\frac{x_{2}}{2}\right)\left(\rho\tau-\frac{1}{4}(\nabla\rho)^{2}\right)+\left(x_{2}+\frac{1}{2}\right)\sum_{q}\left(\rho_{q}\tau_{q}-\frac{1}{4}(\nabla\rho_{q})^{2}\right)\right]\\
&-\frac{1}{16}(t_{1}x_{1}+t_{2}x_{2})J^{2}+\frac{1}{16}(t_{1}-t_{2})\sum_{q}J_{q}^{2}\\
&+\frac{1}{12}t_{3}\rho^{\gamma}\left[\left(1+\frac{x_{3}}{2}\right)\rho^{2}-\left(x_{3}+\frac{1}{2}\right)\sum_{q}\rho_{q}^{2}\right]\\
&+\frac{1}{2}W_{0}\left(J\nabla\rho+\sum_{q}J_{q}\nabla\rho_{q}\right).
\end{split}
\end{equation}
In this expression $q$ represents protons (neutrons) and $\rho$, $J$ and $\tau$ are particle, spin-orbit and kinetic energy densities, respectively.
In the extension of the Hartree-Fock-BCS model to the finite temperature case, temperature dependent Fermi-Dirac
distribution function is used :

\begin{equation}
f_{i}=[1+exp(E_{i}/k_{B}T)]^{-1},
\end{equation}
where $E_{i}$ is quasiparticle energy, $k_{B}$ is the Boltzmann
constant and T is the temperature. In the finite temperature case, BCS equations keep the same structure. However, the occupation factor $v_{i}^{2}$ is changed by
\begin{equation}
v_{i}^{2}(1-f_{i})+u_{i}^{2}f_{i}
\end{equation}
which also modifies the normal
(particle, spin and kinetic energy) and  abnormal densities. Such thermal averaged particle densities are given by \cite{go81,san04},

\begin{equation}
\rho_{T}(\textbf{r})=\frac{1}{4\pi}\sum_{i}(2j_{i}+1)\left[v_{i}^{2}(1-f_{i})+u_{i}^{2}f_{i}\right]\left|\phi_{i}(\textbf{r})\right|^{2},
\label{99}
\end{equation}
\begin{equation}\begin{split}
\textit{J}_{T}(\textbf{r})=&\frac{1}{4\pi}\sum_{i}(2j_{i}+1)\left[j_{i}(j_{i}+1)-l_{i}(l_{i}+1)-\frac{3}{4}\right]\\
&\times\left\{v_{i}^{2}(1-f_{i})+u_{i}^{2}f_{i}\right\}\left|\phi_{i}(\textbf{r})\right|^{2},
\end{split}
\end{equation}
\begin{equation}\begin{split}
\tau_{T}(\textbf{r})=&\frac{1}{4\pi}\sum_{i}(2j_{i}+1)\left[v_{i}^{2}(1-f_{i})+u_{i}^{2}f_{i}\right]\\
&\times\left[\left(\frac{d\phi_{i}(\textbf{r})}{dr}-\frac{\phi_{i}(\textbf{r})}{r}\right)^{2}+\frac{l_{i}(l_{i}+1)}{r^{2}}\left|\phi_{i}(\textbf{r})\right|^{2}\right]
\end{split}
\end{equation}
\begin{equation}
\kappa_{T}(\textbf{r})=-\frac{1}{4\pi}\sum_{i}(2j_{i}+1)u_{i}v_{i}(1-2f_{i})\left|\phi_{i}(\textbf{r})\right|^{2},
\end{equation}
where $v_{i}$ and $u_{i}$ are the BCS variational parameters and $\phi_{i}(\textbf{r})$ the single particle wave function. 
The Finite temperature Hartree-Fock BCS equations for the Skyrme interaction are obtained with the variation of the expectation value of $H^{'}$:

\begin{equation}
\delta\left\langle BCS\left|H'\right|BCS\right\rangle=0.
\end{equation}
In coordinate space, Hartree Fock equation is written as
\begin{equation}
\left\{-\nabla\frac{\hbar^{2}}{2m^{*}\left(\textbf{r}\right)}\nabla+U\left(\textbf{r}\right)+W\frac{1}{i}\left(\nabla\times\sigma\right)\right\}\phi_{i}\left(\textbf{r}\right)=\varepsilon_{i}\phi_{i}\left(\textbf{r}\right).
\end{equation}

In our calculations, the pairing correlations are taken into account with a zero-range density-dependent pairing interaction \cite{ber91}
\begin{equation}\begin{split}
V_{pair}(\textbf{r}_{1},\textbf{r}_{2})&=V_{0}\left(1-\frac{\rho(\textbf{r})}{\rho_{0}}\right)\delta(\textbf{r}_{1}-\textbf{r}_{2})\\
&=V_{eff}(\rho(\textbf{r}))\delta(\textbf{r}_{1}-\textbf{r}_{2}),
\end{split}
\end{equation}
where $V_{0}$ is the pairing strength, $\rho(\textbf{r})$ is the particle density and $\rho_{0}=0.16$ $fm^{-3}$ is the nuclear saturation density. All calculations are carried out with the surface-type pairing interaction. In order to provide a suitable pairing interaction, the pairing strength $V_{0}$ is fixed for each nuclei according to the $\Delta=12/\sqrt{A}$ MeV phenomenological rule. The pairing field is given by
\begin{equation}
\Delta_{T}(\textbf{r})=\frac{V_{eff}(\rho(\textbf{r}))}{2}\kappa_{T}(\textbf{r}).
\end{equation}
In addition, the particle number and pairing energy are also impacted
with the temperature dependent Fermi-Dirac function. The particle
number is defined as
\begin{equation}
N_{q,T}=\sum_{i}(2j_{i}+1)\left[v_{i}^{2}(1-f_{i})+u_{i}^{2}f_{i}\right]
\end{equation}
and the pairing energy is
\begin{equation}
E_{pair,T}=-\frac{\Delta_{T}(\textbf{r})}{2}u_{i}v_{i}(2j+1).
\end{equation}

In the following study, it is relevant to derive the effective nucleon mass $(m^{*}_{q})$ in terms of the nucleon densities \cite{doba84}
\begin{equation}\begin{split}
\frac{\hbar^{2}}{2m^{*}_{q}}=&\frac{\hbar^{2}}{2m}+\frac{1}{4}\left[t_{1}\left(1+\frac{x_{1}}{2}\right)+t_{2}\left(1+\frac{x_{2}}{2}\right)\right]\rho\\
&+\frac{1}{8}\left[t_{2}\left(1+2x_{2}\right)-t_{1}\left(1+2x_{1}\right)\right]\rho_{q}.
\label{112}
\end{split}
\end{equation}

All calculations are performed in a 20 fm box with the assumption of spherical symmetry. The cut-off energy is determined with certain number of states around the Fermi energy. In our calculations, the pairing window (the number of levels above the Fermi level) is taken as 20 for both protons and neutrons. The energy difference between the Fermi and the last considered level is around 15 MeV which is large enough to take care of the occupation of levels due to temperature effects. 
\section{Results}

In this work, we performed FT-HFBCS calculations using zero-range
Skyrme forces. Average neutron pairing gap is calculated via \cite{mat01}:

\begin{equation}
\left\langle \Delta_{n}\right\rangle_{\kappa}=\frac{\int d\textbf{r}\kappa_{T}(\textbf{r})\Delta_{T,n}(\textbf{r})}{\int d\textbf{r}\kappa_{T}(\textbf{r})}.
\end{equation}

Calculations are performed for temperatures up to 1.5 MeV to avoid
occupation of unbound single particle levels and additional
contributions from neutron vapour \cite{bon84}.

\subsection{Tin isotopes}

\begin{figure}[H]
	\includegraphics[width=1. \textwidth]{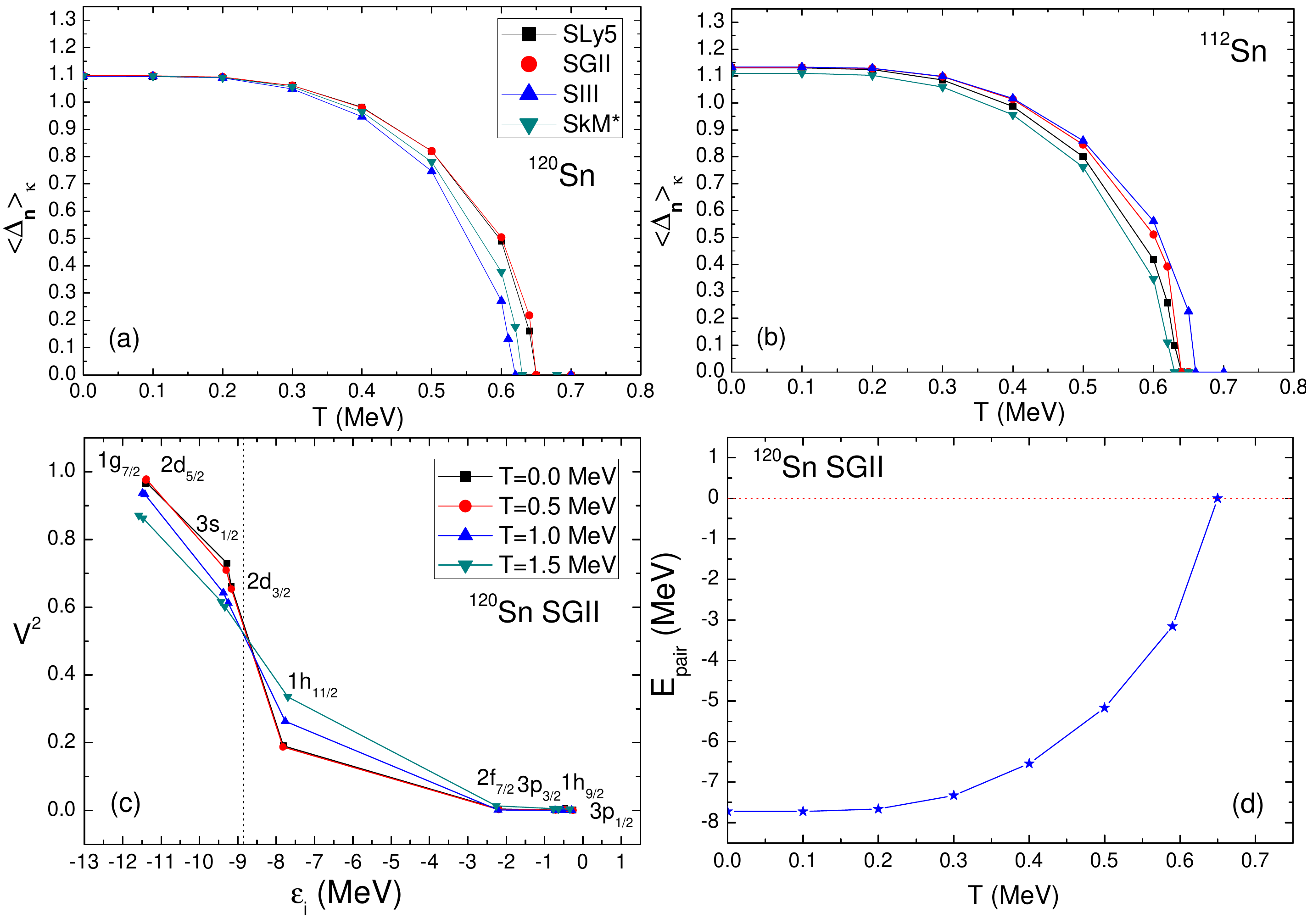}
		\caption{Mean value of the neutron pairing gap as a function of the temperature for $^{120}$Sn (a) and $^{112}$Sn nuclei (b), occupation probabilities for neutrons as a function of temperature around the Fermi level (c). In $^{120}$Sn, the dotted line represents the Fermi level at T=$0.0$ MeV. Pairing energy as a function of temperature (d).}	
		\label{120}
	\end{figure}

In panels (a) and (b) of Figure \ref{120}, the average neutron pairing gap
is plotted as a function of temperature for $^{120}$Sn and $^{112}$Sn
nuclei using various Skyrme interactions. The critical temperature
for $^{120}$Sn and $^{112}$Sn nuclei are found at around 0.65 MeV. With the increase of the temperature, pairing
correlations are destroyed below the Fermi level. As a result of the
high temperature, Cooper pairs are broken and particles are excited to the
higher energy levels which leads to the disappearance of pairing
correlations \cite{egi00}. 

For the Skyrme interactions used (SLy5  \cite{SLy5}, SGII  \cite{SGII}, SIII  \cite{SIII}, SkM*  \cite{SkM}), the critical temperature and the
pairing gap value at zero temperature follow the $T_{c}\simeq
0.57\Delta_{T=0}$ rule. This is also consistent with the finite temperature HFB
results in Ref. \cite{khan07} in which the critical temperature was
found around 0.7 MeV for the $^{124}$Sn nucleus with SLy4 interaction. Calculations are
also performed in A=106 to A=124 even-even Tin isotopes and similar
results are obtained.

In order to provide a clear explanation for the vanishing pairing
mechanism, in panel (c) of Fig. \ref{120} we display the
temperature dependence of the occupation probabilities with respect to
the single particle energies of the $^{120}$Sn nucleus using the SGII
interaction. The single particle levels are chosen around the Fermi
level, which are impacted by temperature. With the increase of the temperature, nucleons are excited to higher energies and occupation
probabilities are increased above the Fermi level. Especially, the
occupation probability of the $1h_{11/2}$ state just above the Fermi
level increases above the critical temperature while the occupation probabilities of $3s_{1/2}$ and $2d_{3/2}$ 
states are decreasing below the Fermi level. In addition, the change of the pairing
energy as a function of temperature is displayed in panel (d) of Fig. \ref{120}.
Although the decrease of the pairing energy is small up to 0.3 MeV, a substantial decrease
of the pairing energy is obtained for the temperatures between 0.3 MeV and 0.6 MeV. At
critical temperature, the pairing energy is exactly zero, showing
the vanishing of pairing correlations and interpreted as the phase
change from superfluid to the normal state. The sharp
transition is due to the limitation of the present approach which corresponds
the grand-canonical ensemble \cite{egi93}. Similar
results are also obtained with other Tin isotopes and Skyrme
interactions: SLy5, SIII and SkM*. 

\begin{figure}[H]
	\includegraphics[width=1. \textwidth]{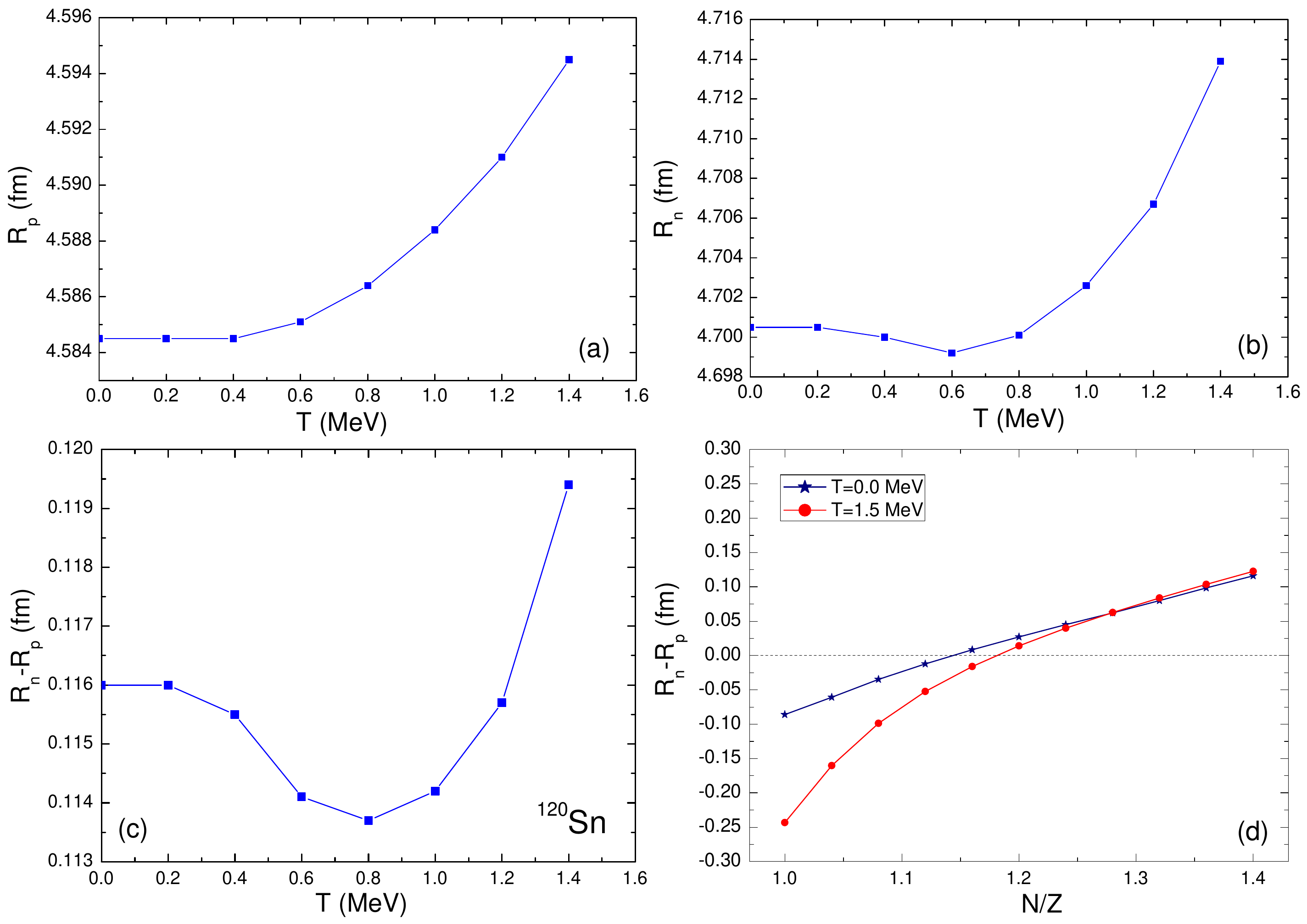}
		\caption{Proton (a) and neutron (b) radius with respect to the temperature for $^{120}$Sn, calculated with SGII interaction and the change of neutron skin radius of $^{120}$Sn nucleus with respect to the temperature (c). $R_{n}-R_{p}$ difference as a function of N/Z value for the Sn isotopes at T=0.0 MeV and T=1.5 MeV (d).}
		\label{122}
	\end{figure}

In panels (a) and (b) of Fig. \ref{122}, the change of the
proton and neutron radii with respect to the temperature is displayed for the
$^{120}$Sn nucleus with SGII interaction. The interaction SGII \cite{SGII} is one of the widely
used parameterizations of the Skyrme force. It has been derived from SkM*,
by improving the Landau parameters and in particular $G_{0}^{'}$. This ensures
a good description of spin states like the Gamow-Teller resonance, in
addition to the reproduction of ground state masses and radii.
The proton radius stays constant
up to the critical temperature and then increases linearly. Although a
small decrease is obtained in the case of the neutron radius up to the critical
temperature, it also increases after critical temperature.  The
temperature dependence of the neutron skin ($R_{n}-R_{p}$) is also presented in the
panels (c) and (d) of Fig. \ref{122}. In the case of $^{120}$Sn
although a small decrease is obtained before the critical temperature,
the neutron skin increases of about 3\% at temperatures above 1 MeV.
This is due to the increasing population of larger single-particle
states with larger angular momentum (\textit{l}), such as the $1h_{11/2}$ state, as discussed above. Investigation of the
effect of N/Z value on the temperature dependence of the neutron skin
is also performed in the Tin isotopic chain. In panel (d) of Fig.
\ref{122}, the $R_{n}-R_{p}$ difference with respect to N/Z value is
displayed. It has been known that HFBCS is not suitable for drip line nuclei due to the formation of unphysical neutron gas \cite{dob96}. 
Therefore, we limit our calculations up to  $^{120}$Sn nucleus in this part.
 Since both proton and neutron radii increase above the
critical temperature, the proton radius weakens the formation of
neutron skin in less neutron-rich Tin isotopes with the increase of the temperature.

\begin{figure}[H]
	\includegraphics[width=1. \textwidth]{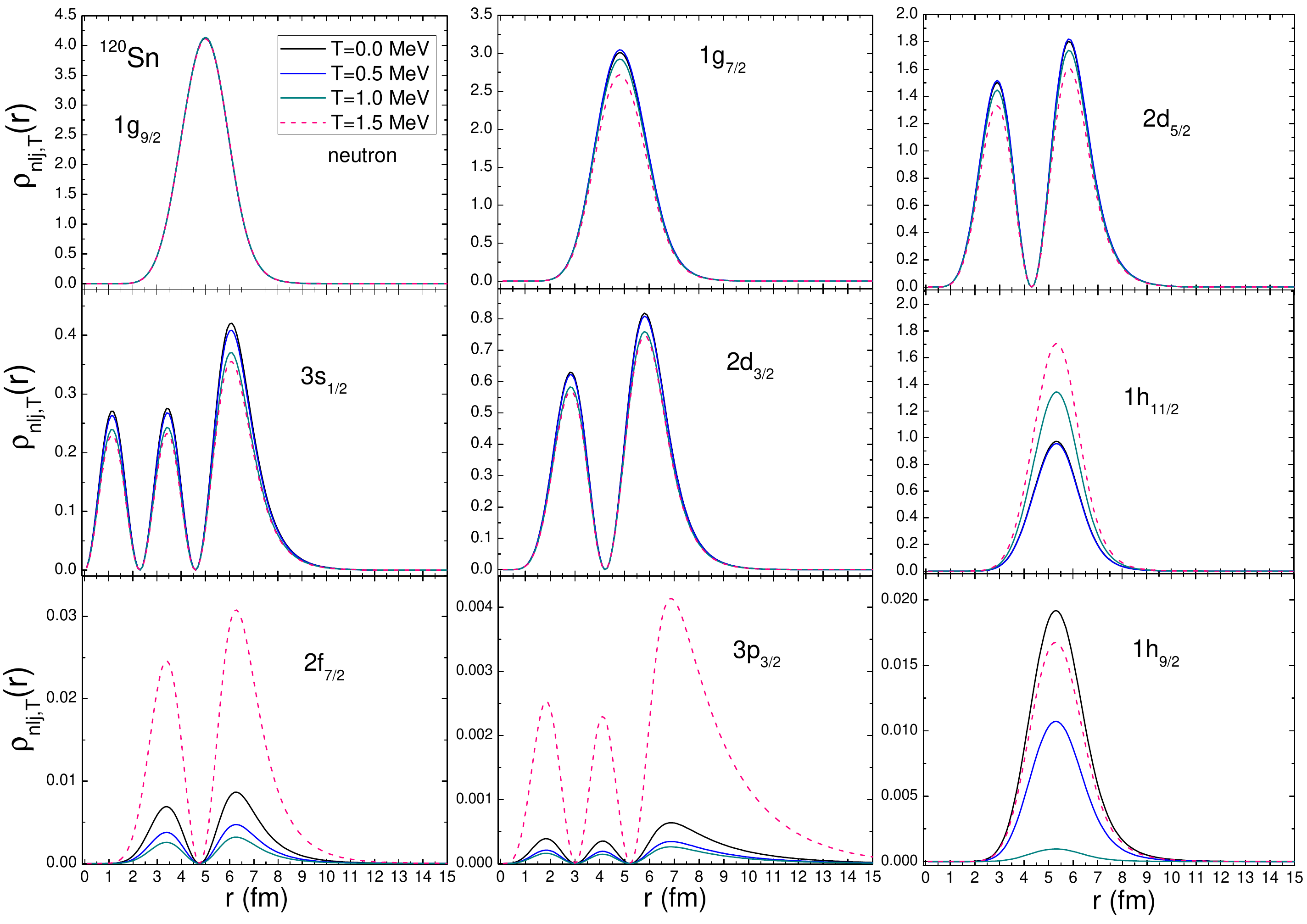}
		\caption{Neutron partial densities of states (see Eq. 
\eqref{99}) as a function of temperature for $^{120}$Sn nucleus with SGII interaction. Here, $n, l, j$ are representing the principal, orbital angular momentum and total angular momentum quantum numbers of the corresponding state, respectively.}	
		\label{123}
	\end{figure}

In order to further investigate the increase of the neutron skin with
temperature, in Fig. \ref{123} the neutron partial densities
of states ($\rho_{nlj,T}(r)$) as a function of temperature are displayed for the $^{120}$Sn nucleus.
Since temperature impacts mostly the states around the Fermi level, the single
neutron levels around the Fermi level are displayed. First, the effect of
temperature is seen in the $1g_{7/2}$ single particle level which is below
the Fermi level. While the neutron partial densities are decreasing
below the Fermi level, an increase in the partial density of the
$1h_{11/2}$ state just above the Fermi level is obtained at high
temperatures. Proton partial densities also show a similar behavior:
while partial densities decrease below Fermi level, an increase is
obtained above the Fermi level at high temperatures. This causes an
increase in the proton radius. In the case of 
the $2f_{7/2}$, $3p_{3/2}$ and $1h_{9/2}$ neutron states, a more
subtle effect is at work: the partial densities decrease up to T=1.0 MeV 
and then increase. The decrease is due to the temperature effects
which are destroying the small occupation number generated by
pairing effects. Above 1 MeV, these states start to be repopulated by
temperature effects, because of their increasing Fermi-Dirac factors
(feeding from lower energy states).

\subsection{Temperature effects on effective masses}	
	
\subsubsection [Effective mass of symmetric nuclear matter and
\texorpdfstring{$(T_{c}$}{(Tc)}]      {Effective mass of symmetric
nuclear matter and $T_{c}$}

In this section, our aim is to investigate the possible relation
between the effective mass in symmetric nuclear matter (or
equivalently m*(r=0)) and the critical temperature. 
The effective mass is known to impact the level density
and pairing correlations which in turn may have an effect on the
critical temperature \cite{vaut72,mah85}. In our calculations,
${m^{*}}/{m}$ values are ranging from 0.6 to 1.0 for the selected
Skyrme energy density functionals.

In symmetric nuclear matter, effective masses for the
Skyrme energy density functionals are defined as \cite{SLy5}:

\begin{equation}
\frac{m^{*}_{Sky.}}{m}=\left[1+\frac{m\rho_{0}}{8\hbar^{2}}\left\{3t_{1}+t_{2}\left(5+4x_{2}\right)\right\}\right]^{-1},
\label{111}
\end{equation}
where $\rho_{0}$ is the nuclear saturation density.

\begin{figure}[!htb]
	\includegraphics[width=0.95 \textwidth]{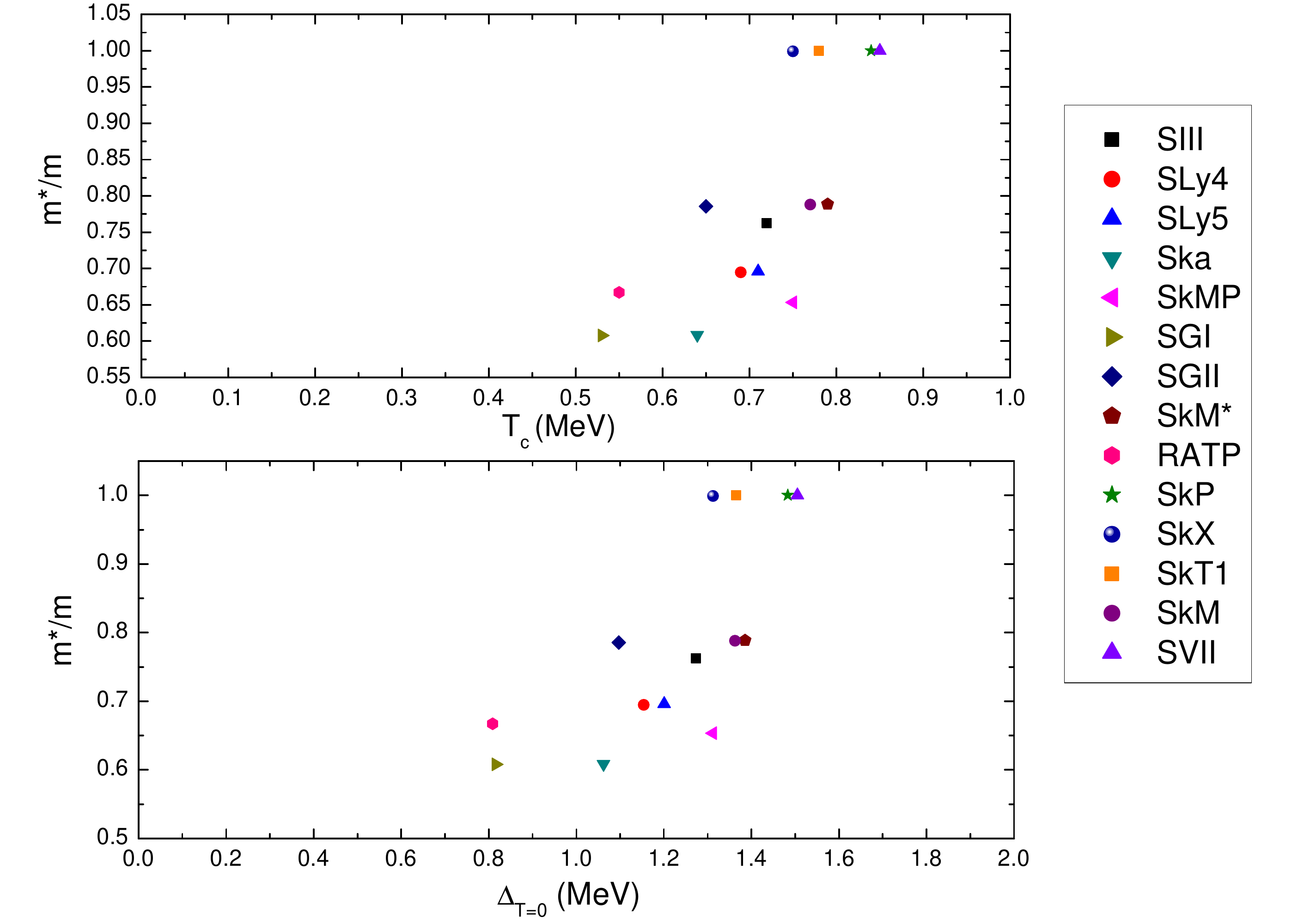}
		\caption{Upper Panel: Effective mass values with respect to the critical temperature. Lower Panel: Effective mass values with respect to the mean value of the pairing gap. Calculations are performed for $^{120}$Sn nucleus using 14 different Skyrme interactions.}	
		\label{150}
	\end{figure}

\begin{figure}[H]
	\includegraphics[width=0.95 \textwidth]{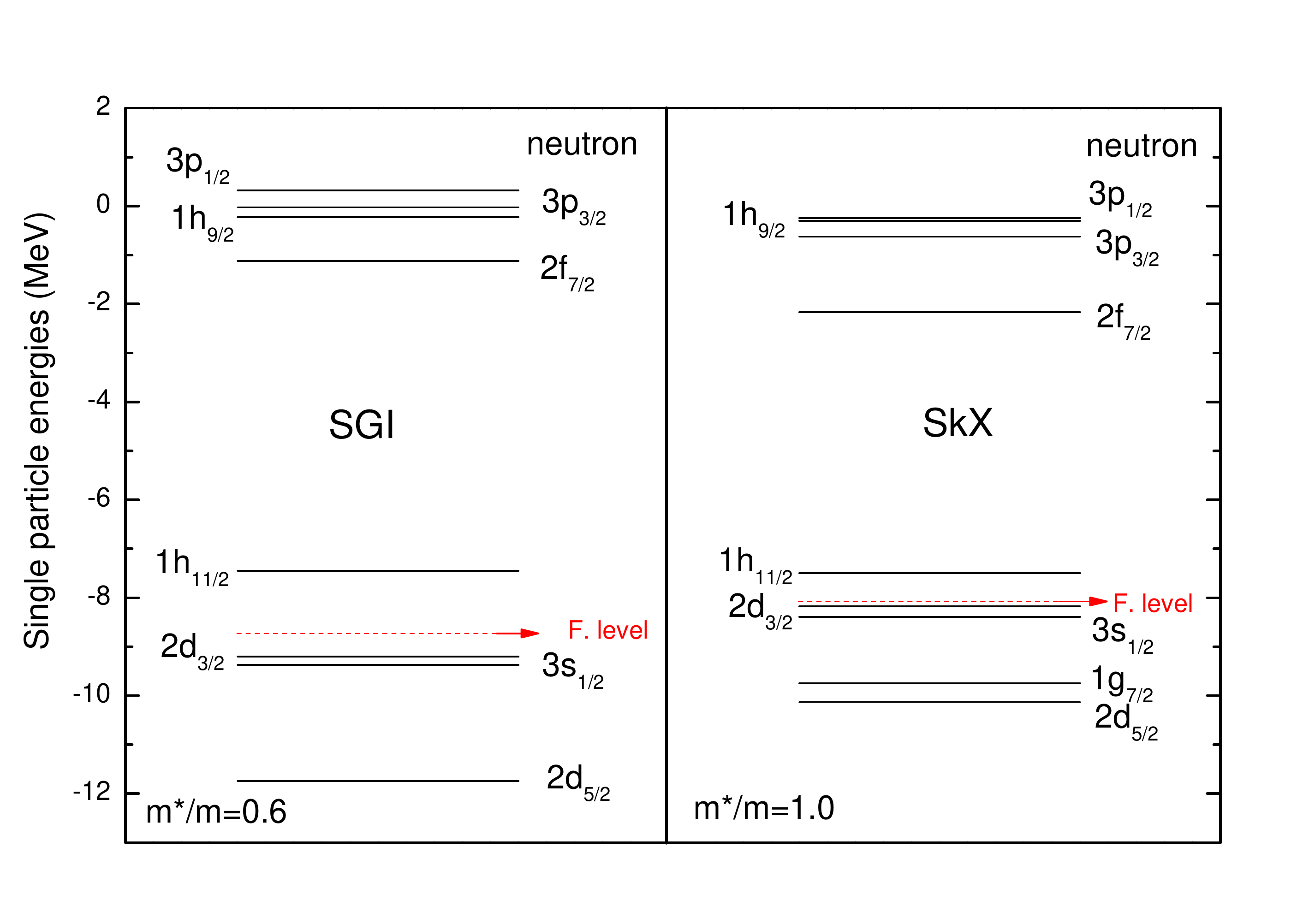}
		\caption{Neutron single particle energies around the Fermi level for $^{120}$Sn nucleus with SGI and SkX interactions, at zero temperature.}	
		\label{152}
	\end{figure}

The upper panel of Figure \ref{150} shows the influence of the
effective mass values of the Skyrme energy density functionals (Eq.\eqref{111}) on the critical temperature of $^{120}$Sn nucleus.
Calculations are performed with 14 different Skyrme interactions, keeping the pairing strength constant, $V_{0}=680$ $MeV.fm^{3}$, for all interactions used. A
correlation between the effective mass and the critical temperature is
obtained, namely with the increase in the effective mass values of the
Skyrme interactions the critical temperature increases. Since a larger
effective mass generates larger level densities and larger pairing gap
values (see lower panel of Fig. \ref{150}) , the critical temperature at which the pairing correlations are
disappearing is also larger. The pairing gap at zero temperature
and the critical temperature are strongly correlated by the
$T_{c}\simeq 0.57\Delta_{T=0}$ law and larger effective mass value
generates a higher pairing gap value as expected. To check this interpretation, we also plot neutron single particle
energies for the $^{120}$Sn nucleus with SGI and SkX Skyrme interactions
(see Fig. \ref{152}) which have 0.6 and 1.0 ${m^{*}}/{m}$ values,
respectively. SGI interaction has lower effective mass value and
predicts larger energy difference between the $2d_{3/2}$ level and
Fermi level (smaller level density).

This low level density around the Fermi level gives smaller pairing
gap value. Therefore the critical temperature value for the phase
transition is lowered.  The SkX interaction has a larger effective
mass and predicts smaller energy difference between the $2d_{3/2}$
level and Fermi level. Hence the level densities and pairing gap are
enhanced which eventually causes a larger critical temperature value. We
checked that same calculations with different Skyrme interactions give
similar result. 

It should be noted that in the case of imposing a constant pairing gap, the critical temperature is also constant due to the above relation between the pairing gap and the critical temperature. In this case the critical temperature dependence of the 
effective mass vanishes.

\subsubsection{Temperature dependence of the effective mass in nuclei}

In this section, we investigate the effect of temperature on the
effective masses of nuclei. The total effective nucleon mass is
obtained with the product of the effective k-mass and $\omega$-mass
$(m^{*}/m=m_{\omega}m_{k}/m^{2})$. While the k-mass comes from the spatial
non-localities and is obtained from the mean field calculations,
the $\omega$-mass is related to the frequency dependence of the mean field
\cite{fan10}.

The level densities around the Fermi level are
affected with the changes in the effective mass as discussed above. Therefore the
determination of the temperature dependence of the effective mass is
an important task for the level density calculations. The temperature
dependence of the effective nucleon mass of nuclei have been
previously studied in Refs. \cite{hasse86,bor87,fan10,don94}. The
results indicate a decrease of the effective nucleon masses at high
temperatures. Therefore the nuclear symmetry energy and electron
capture rates in the stellar core collapse, which takes place at 1-2
MeV temperature are also affected \cite{fan10}. In ref.\cite{don94}, the
temperature dependence of the effective $\omega$-mass was investigated with
respect to the temperature. The effective mass was found to decrease which
decreases the level density and increases symmetry energy
at high temperatures. The changes in the
effective proton and neutron masses might have some effects on
astrophysical events; for example in the electron capture rates of
stellar collapses \cite{don94}. Since the rate of the electron capture
is closely related with the proton and neutron chemical potentials and
consequently with the symmetry energy, an increase or decrease in the
effective nucleon masses might lead significant changes in these
rates. 

\begin{figure}[!hbt]
	\includegraphics[width=1. \textwidth]{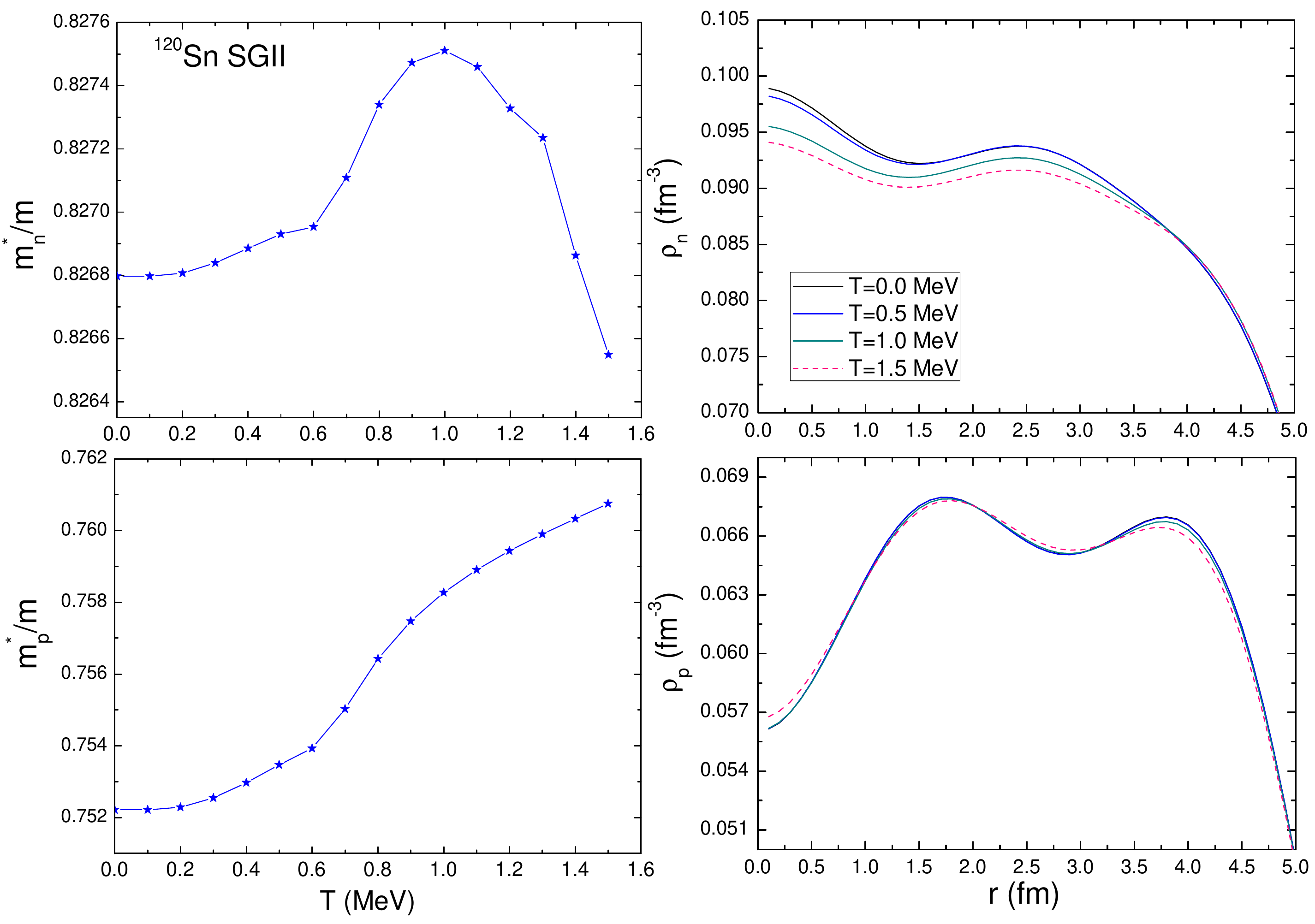}
		\caption{Left Panel: Effect of temperature on the
		neutron and proton effective masses (at r=0) for $^{120}$Sn nucleus with SGII interaction. Right Panel: proton and neutron densities with respect to the temperature.}	
		\label{151}
	\end{figure}

In this framework, in the left part of Fig. \ref{151}, the effect of temperature on the
effective k-masses (at r=0, Eq.\eqref{112}) of protons and neutrons are
displayed. The results are shown for $^{120}$Sn nucleus with SGII
interaction. 

An increase of the effective neutron mass up to T=1.0 MeV is observed
in the interior of the nuclei. Thereafter, the effective neutron mass
suddenly decreases from T=1.0 MeV to
T=1.5 MeV. In addition, the increase of the proton
effective mass is much more pronounced after the critical
temperature. The change is about $ 1.13\%$ for proton effective mass. Although the changes in the effective masses are small in magnitude as mentioned in previous works \cite{fan10}, it may be relevant to accurately study its behavior.
To explain the behavior of the effective proton and neutron masses, we
rewrite Eq.\eqref{112} using the values of the SGII parameters:

\begin{equation}
\frac{\hbar^{2}}{2m^{*}_{q}}=\frac{\hbar^{2}}{2m}+\left(64.56(MeV.fm^{5})\rho-57.67(MeV.fm^{5})\rho_{q}\right) 
\label{113}
\end{equation}

	\begin{figure}[!hbt]
	\includegraphics[width=1. \textwidth]{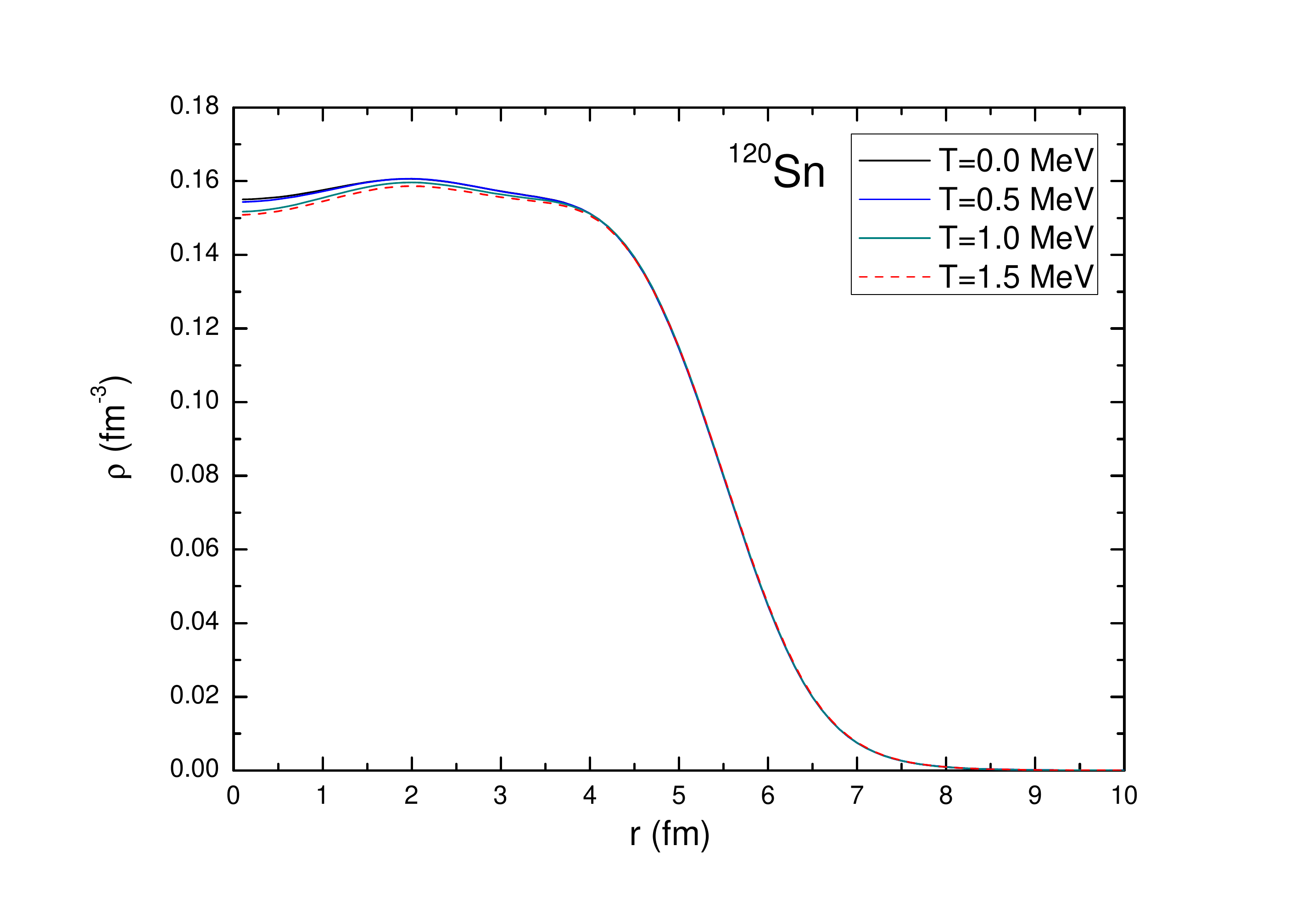}
		\caption{Change of the total density with respect to the 
		temperature in $^{120}$Sn.}	
		\label{totden}
	\end{figure}

The changes in the effective masses depend on the density but not explicitly on the temperature. 
Therefore, the temperature dependence of the effective masses come indirectly from the temperature dependence of the density.
(Eq.\eqref{113}). The total density is displayed in Fig.
\ref{totden}. In the center of the nucleus, the total density decreases 
with increasing temperature. Therefore the effective mass is expected
to increase with increasing temperature, but there is also the effect of
the isospin dependent density in Eq.\eqref{113}: in the right part
of Fig. \ref{151}, we display the change of the neutron and proton
densities with respect to the temperature. The effect of temperature
on the densities are pronounced above T=1.0 MeV i.e., above critical
temperature. After T=1.0 MeV, the neutron density decreases and the
proton density increases in the interior of the nuclei. The decrease
of the neutron density in the interior of the nuclei is directly
linked to by the population of levels above the Fermi level, and
depopulation of levels below (see Fig. \ref{123}), including the neutron
$3s_{1/2}$ state which contributes to the center of the nucleus. As a
result, the neutron (proton) effective mass $(m^{*}_{q}/m)$ at r=0
decreases (increases) after the critical temperature. The present
results provide an accurate view of the temperature
effects on the effective mass in nuclei. In order to provide a general trend for the behavior of the effective k-mass,
 we performed same calculations with other Skyrme interactions: SLy5, SIII and SkM* and also with other nuclei.
 The results reveal that the behavior of the effective k-mass depends on the interaction used and nuclei considered. The k-mass can either increase or decrease, however, the change is small and ranging between $ 0.15\%$ and $ 1.1\%$ from T=0 MeV to T=1.5 MeV. Although in most of the calculations temperature dependence of the k-mass is small and taken as constant at first order, our results show that the temperature dependence of the k-mass should not be ignored in the case of an accurate study of the total effective mass at high temperatures.

\section{Conclusion}

In summary, we have investigated the pairing properties of different nuclei
within the finite temperature HF+BCS framework with Skyrme
interactions. The relation between the critical temperature and the
pairing gap value at zero temperature generally obeys the $T_{c}\cong0.57
\Delta_{T=0}$ equation, depending on the energy difference between the
last occupied and first unoccupied states and the interaction used.
The temperature dependence of the neutron skin is also investigated with
different nuclei. A substantial increase in the proton and neutron radius of
neutron rich nuclei is obtained above the critical temperatures.
However for less-neutron rich nuclei, the increase in the proton radius cancels
out the rapid increase of the neutron skin radius. The changes in the neutron
skin is mainly due to the effect of temperature on the occupation
probabilities of the single-particle states around the Fermi level.
High temperature causes the occupation of the unoccupied levels which
is leading an enhancement in the proton and neutron radius.

A correlation is obtained between the effective mass of
Skyrme energy density functionals and the critical temperature of
$^{120}$Sn nucleus, relating $m^{*}$, T$_c$ and $\Delta(T=0)$. It has been
shown that with the enhancement in the effective mass, the level
densities around the Fermi level and pairing gap increase which
eventually leads an increase in the critical temperature. However this correlation depends on the pairing prescription. Temperature
dependence of the effective k-masses of protons and neutrons are also
investigated. Due to the changes in the proton and neutron densities,
the effective nucleon masses show rather different fine behavior before
and after the critical temperature, which is a new feature.

\end{document}